\documentclass[twocolumn,showpacs,preprintnumbers,amsmath,amssymb]{revtex4}
\usepackage{graphicx}% Include figure files
\usepackage{dcolumn}% Align table columns on decimal point
\usepackage{bm}% bold math

\begin{document}

\preprint{KUCP0216}

\title{ Quasi-normal modes of D3-brane Black Holes}% 

\author{Yasunari Kurita}
 \email{kurita@phys.h.kyoto-u.ac.jp}
\affiliation{%
 Graduate School of Human and Environmental Studies,
 Kyoto University,  Kyoto  606-8501, Japan 
}%

\author{Masa-aki Sakagami}
 \email{sakagami@phys.h.kyoto-u.ac.jp}
\affiliation{
 Department of Fundamental Sciences, FIHS, Kyoto University,
       Kyoto, 606-8501, Japan 
}%

\date{\today}% It is always \today, today,

\begin{abstract}
We investigate a method to evaluate  quasi-normal modes of D3-brane black holes 
by wave interpretation of fields on D3-brane 
based on the Feynman's space-time approach. 
We perturbatively solve the wave equation which describes propagation of a dilaton wave
in a bulk space and its interaction with the D3-brane.
The obtained condition for the quasi-normal modes are qualitatively equivalent to 
that evaluated in the usual scattering of the dilaton in the black 3-brane spacetime 
in the corresponding supergravity description. 
\end{abstract}

\pacs{04.70.-s,04.50.+h,11.25.Hf,11.15.-q}

\maketitle

\section{Introduction}

It is well known that black holes have thermodynamical properties, i.e. 
entropy\cite{entropy,AreaTh} and temperature\cite{HawkingRadiation}, 
which can be understood in the framework of 
general relativity\cite{GRText} and quantum theory of matter 
in curved spacetime\cite{QFText}. 
In these formalisms, the entropy for a black hole is given by 
a quarter of its horizon area and Hawking radiation can be explained as 
a particle creation caused by existence of the horizon.  
Some years ago, in the string theory, which seems to be most promising 
candidate for a quantum theory of gravity,
it was found that the D-brane can also describe 
black hole spacetime and its thermodynamical properties\cite{H}.
In this prescription, the entropy for a black hole 
is derived by counting the number of microscopic states on the D-brane\cite{SV},
and Hawking radiation can be recognized as  emission process of 
closed strings from the D-brane\cite{CM,DM}.

As for the scattering processes of a particle or a wave in   
a black hole spacetime, we can observe a good agreement between a D-brane 
picture and an analysis based on the corresponding supergravity description.  
Especially the absorption cross section for a dilaton by the D3-brane
in a low-energy region,
which is evaluated by means of the world volume approach, 
has been shown to coincide with the result obtained by solving the    
wave equation for the dilaton field propagating in the 3-brane 
background\cite{K,GHKK,GH}.    
Taking account of the above results in the scattering processes, 
it seems to be quite natural to pose a question whether the agreement  
is hold even for quasi-normal modes (QNMs) of a black hole or not.

QNMs characterize the emission of a gravitational 
wave  which represents a response to a perturbation affected to
a black hole spacetime\cite{ChandDet,MathematicalBH,KS}. 
As briefly reviewed in appendix A, it is obtained 
by solving the wave equation on the
 background with the suitable boundary conditions 
that the flux at the horizon is ingoing and 
outgoing at spatial infinity.
QNM is characterized by a complex frequency whose 
imaginary part represents 
the time scale in which the perturbation to the black hole spacetime decays.
Since  black holes are interpreted as  thermal objects
that are  characterized  by thermal quantities such as 
the temperature and entropy, the imaginary part of the QNM frequency  
can be recognized as  the relaxation time within which 
black holes approach to the thermal equilibrium. 

As was well known, the relationship between D-branes and black holes 
in string theory is an important precursor to the AdS/CFT 
correspondence. 
And connection between QNMs and the decay of perturbations
in the dual CFT was first suggested in the work of \cite{HH}
based on the numerical computation of QNMs for AdS-Schwarzschild 
black holes in several dimensions.
QNMs of AdS black holes 
and BTZ black holes are investigated in  
\cite{CM1,CM2,WYA,GS,ZWA,B,CL1,CL2,MN,WAM,RAK} and those of near 
extremal black branes are found in \cite{AOS}.
Furthermore in \cite{BSS}, it was shown that  the frequencies of 
QNMs for BTZ black holes 
 are in exact agreement with the location of the poles of the 
retarded correlation function describing the linear response 
 on the CFT side.

In the present paper, we consider the D3-brane 
as an other example of the correspondence in QNMs.
Our analyses for  the D3-branes will be performed in the 
two different parameter regions in type IIB string theory, 
one is large number of the  D3-branes 
and the other is only one D3-brane.
In the first region, the classical supergravity is effective so that  
gravity is described by the curvature of space-time.
In the later, the perturbative field theory on the D3-brane, 
which is embedded in a flat spacetime,  is effective.  
In such flat spacetime, how QNMs can be calculated?
This is our main interest and motivation in this paper. 
We will apply Feynman's space-time approach to this problem,
which gives alternative intuitive methods to quantum field theories
\cite{Feynman 1,Feynman 2,Feynman 3,JJS}.
We perturbatively solve the equation derived from 
the low energy effective action for the D3-brane,
i.e. the Dirac-Born-Infeld action, which describes propagation 
of a dilaton wave in the bulk flat spacetime and its interaction with
D-branes. And we apply a general condition for QNMs which is 
obtained in appendix A to the scattering problem of this flat D3-brane case.
It is shown that the obtained condition of QNMs is qualitatively equivalent 
to that evaluated in the framework of the usual scattering of 
the dilaton wave based of the supergravity description.  

A brief outline of this paper is following.
In the next section, we obtain the QNMs for the 3-brane by solving 
the wave equation propagating on its background 
in classical supergravity analysis.
In section 3, only one D3-brane are considered and 
the condition of the QNMs are obtained by the Feynman's space-time approach.
These two results are compared and discussed in section 4.  
In appendix A, we give a brief review of QNMs.
Absorption of a scalar for D3-brane is also investigated 
by space-time approach and the result is compared with the result of 
\cite{GHKK} in appendix C.

\section{Quasi-normal modes in Supergravity}

In this section, we evaluate the QNM of 3-brane solution of type IIB
supergravity.
The main analysis of this section is based on the results obtained in 
{\cite{GH}}, which studies
absorption probability of a dilaton by the D3-brane.
We consider the dilaton as a minimally coupled scalar, 
which obeys the wave equation
%----------------------Eqn of Dilaton--------------------------------------------%
\begin{equation}
\frac{1}{\sqrt{-g}} \partial_A \sqrt{-g} 
  g^{A B} \partial_B \phi = 0 , \quad (A,B=0,1,\dots,9)
\label{min-eq}
\end{equation}
%--------------------------------------------------------------------------------%
in the spacetime whose  metric is given by 
%----------------------3-brane metric in type IIB SUGRA-------------------------%
\begin{eqnarray}
ds^2 = && \left(1+\frac{R^4}{r^4}\right)^{-1/2} (-dt^2 + dx_i dx^i) \nonumber \\
&& +\left( 1+\frac{R^4}{r^4} \right)^{1/2} (dr^2+r^2d {\Omega}_5^2). 
\end{eqnarray}
%_______________________________________________________________________________%
The characteristic length of the 3-brane  $R$ is related to 
10-dimensional gravitational
 coupling constant  $\kappa_{10}$ as follows,
%---------------------Relation of RR charge--------------------------------------%
\begin{equation}
  R^4 = 4 \pi g N {\alpha'}^2 = \frac{N \kappa_{10}}{2 {\pi}^{5/2}}.
\end{equation}
%--------------------------------------------------------------------------------%
In the case of the scattering in the low energy region
$\omega R << 1$,
the dominant contribution to the cross section comes from 
spherical symmetric process so that
we concentrate   
the radial equation for s-wave of energy $\omega$ derived from
Eq.(\ref{min-eq})  
%--------------------------Equation of motion for s-wave--------------------------%
\begin{equation}
\left[{\partial^2 \over \partial r^2} + {5 \over r} {\partial
\over \partial r} + \omega^2 \left(1+ {R^4
\over r^4} \right) \right] \phi(r) = 0 . 
\label{s-wave}
\end{equation}
%--------------------------------------------------------------------------------%
If one performs the change of variables, 
$r = R e^{-z}$, $\phi(r) = e^{2z} \psi(z)$,  
then Eq.(\ref{s-wave}) becomes Mathiue equation,
%---------------------------------Mathiue equation------------------------------%
\begin{equation}
\left[ {\partial^2 \over \partial z^2} + 
 2 (\omega R)^2 \cosh 2z - 4 \right] \psi(z) = 0.
\label{maineq2}
\end{equation}
%--------------------------------------------------------------------------------%
As was shown in \cite{GH},
the exact solution which is ingoing at horizon($r \to 0$) can be expressed 
as expansions in terms of  Bessel and H$\ddot{a}$nkel functions as follows,
%--------------------------------Exact solution----------------------------------%
\begin{equation} 
 \psi(z) = \sum_{n=-\infty}^\infty 
 {c\left( n + \mu \right) \over 
 c(\mu)} J_n(\omega R e^{-z})
 H^{(1)}_{n+2\mu}(\omega R e^z) ,
\label{sol}
\end{equation}
%--------------------------------------------------------------------------------%
where the coefficients is   given by
%-------------------------------Floquet exponent---------------------------------%
\begin{eqnarray}
 c(\mu) &=& {({\omega R \over 2})^{2\mu} \over \Gamma(\mu+2) \Gamma(\mu)} v(\mu) \nonumber \\
 v(\mu) &=& \sum_{n=0}^\infty (-1)^n \left({\omega R \over 2}\right)^{4n}
               A_{\mu}^{(n)} \nonumber \\
 A_{\mu}^{(0)} &=& 1 \nonumber \\ 
 A_{\mu}^{(q)} &=& \sum_{p_1=0}^\infty \sum_{p_2=2}^\infty \ldots
\sum_{p_q=2}^\infty a_{\mu+p_1} a_{\mu+p_1+p_2} \cdots
a_{\mu+p_1+\ldots+p_q} \nonumber \\
 a_{\mu} &=& {1 \over \mu(\mu+1)(\mu+2)(\mu+3)}.
\label{Floquetcoef}
\end{eqnarray}
%--------------------------------------------------------------------------------%
The value of $\mu$ is determined in terms of a prescription in 
the standard Floquet analysis which implies 
%------------------------------------------------------------------------------%
\begin{equation}
\frac{c(\mu)}{c(\mu -1)}\frac{c(-\mu+1)}{c(-\mu)} =1 .
\label{Floquet}
\end{equation}
%-------------------------------------------------------------------------------%
The explicit expression for the first few terms of $\mu$ and $A_{\mu}^{(q)}$ is
given in Appendix B.
Let me introduce new variables for convenience,
%-------------------------------------eta-xi-------------------------------------%
\begin{equation}
   \eta = e^{2i\pi\mu}, \qquad \chi = {c(-\mu) \over c(\mu)}. 
\label{eta-xi}
\end{equation}
%---------------------------------------------------------------------------------%
With these variables, asymptotic forms of the solution (\ref{sol}) near 
the horizon ($\rm{Re}~ z \to \infty$) is given by
%-----------------------------Asymptotic behavior---------------------------------%
\begin{eqnarray}
   \sqrt{\eta}& \displaystyle \left( \eta - {1 \over \eta} \right) \psi(z)&   
\nonumber \\
    \to& \displaystyle
       ~~~~~\left( \eta - {1 \over \eta} \right) &
         \sqrt{2 \over \pi \omega R e^z} 
         e^{i \left( \omega R e^z - {\pi \over 4}  \right)}.  
\end{eqnarray}
%-------------------------------------------------------------------------%
Similarly we can obtain  the asymptotic form for the spatial infinity
(${\rm Re}~ z \to -\infty$)  
%--------------------------------------------------------------------------%
\begin{eqnarray}
 \sqrt{\eta}& \displaystyle \left( \eta - {1 \over \eta} \right) \psi(z)& 
\nonumber \\
 \to & \displaystyle
  ~~~~~\left( \chi - {1 \over \chi} \right) &
         \sqrt{2 \over \pi \omega R e^{-z}} 
         e^{i \left( \omega R e^{-z} - {\pi \over 4} \right)}  \nonumber \\
      & \displaystyle + \left( \eta \chi - 
        {1 \over \eta \chi} \right) &
         \sqrt{2 \over \pi \omega R e^{-z}} 
         e^{-i \left( \omega R e^{-z} - {\pi \over 4} \right)}.\nonumber \\
\label{asympt}
\end{eqnarray}
%----------------------------------------------------------------------------------%
From these asymptotic behavior given above, we can read off the amplitudes 
$${\mathcal{A}}=( \eta - {1 \over
\eta}), ~~~~{\mathcal{I}}=(\eta\chi - {1 \over \eta\chi}), 
~~~~{\mathcal{R}} =(\chi-{1 \over \chi}),
$$
 for the transmitted, incident and reflected waves, respectively.
As we describe in Appendix A, QNMs are  given 
by the condition,
%------------------------------The general condition of QNM---------------------%
\begin{equation}
{\mathcal{I}}/{\mathcal{R}} = 0.
\label{cond.QNM}
\end{equation}
%-------------------------------------------------------------------------------%

In order to compare the obtained results (\ref{cond.QNM}) with 
the evaluation of QNMs in the next section based on Dirac-Born-Infeld 
action which is the low energy effective theory of the D3-branes,
we consider the QNM condition (\ref{cond.QNM}) 
in the low energy region ,i.e.,  $\omega R \ll 1$, 
that can be expanded with respect to $\omega R$ as 
%---------------------------expansion---------------------------------%
\begin{widetext}
\begin{eqnarray}
{{\mathcal{I}} \over {\mathcal{R}}}e^{i\beta}
& =& 1+\frac{\pi^2}{512}(\omega R)^8
\left(1+\frac{4i}{\pi}\log\frac{\omega \bar{R}}{2} \right) 
+ \frac{\pi^2}{512\cdot 12} (\omega R)^{12}
\left(
-2\log\frac{\omega \bar{R}}{2}+\frac{4i}{\pi}
(\log\frac{\omega \bar{R}}{2})^2
\right) \nonumber \\
 & & + \frac{\pi^2}{512}\frac{7}{72}(\omega R)^{12}
\left(
1+\frac{4i}{\pi}\log\frac{\omega \bar{R}}{2}
\right)+{\mathcal{O}}\left(
(\omega R)^{16}(\log\omega R)^3
\right) \label{eq for QNM in SUGRA}\\
&& \nonumber \\ 
&=& 0, \nonumber
\end{eqnarray}
\end{widetext}
%---------------------------------------------------------------------------------%
where $\bar{R}$ is written by the characteristic length of the 3-brane $R$ 
and the Euler constant $\gamma$ as  $\bar{R}= e^{\gamma}R $ and
$\beta$ is a phase factor,
%----------------------Phase factor----------------------------------------------%
\begin{equation}
\beta = \frac{2}{3}\pi(\frac{\omega R}{2})^4+\frac{259}{216}\pi
(\frac{\omega R}{2})^{8}-\frac{22}{81}\pi^3(\frac{\omega R}{2})^{12}+\cdots
\end{equation}
%-------------------------------------------------------------------------------%

Let us solve approximately 
the above QNM condition up to the lowest order $(\omega R)^8$ , 
%---------------------------------eq. for QNM to order 8---------------------------%
\begin{equation}  
1+\frac{\pi^2}{512}(\omega R)^8
\left(1+\frac{4i}{\pi}\log\frac{\omega \bar{R}}{2} \right) 
= 0.
\label{eq.for QNM to 8}
\end{equation}
%---------------------------------------------------------------------------------%
In the usual evaluation to QNMs \cite{WYA,GS,ZWA,B,CL1,CL2,MN,RAK},
their frequencies are characterized by the curvature scale of the black
hole space time, i.e. $|\omega R|\sim 1$.
Contrastingly, our calculation, which is valid in the range to  
$|\omega R|\ll 1$, gives QNMs in the low energy region.
From the observational point of view,
the frequencies for QNMs obtained in this paper might not be important.
However recall that our main interest is to confirm the equivalence
between two different pictures, i.e. the D3-branes and black 3-branes, in
the case of QNMs.

From the condition (\ref{eq.for QNM to 8}) in the region 
$|\omega R|\ll 1$, we note that its solutions lie on n-th Riemanian sheets
with $n\gg 1$ in the complex $\omega R$ plane.
Inserting  a polar coordinate representation for
$(\omega R)^8$ to Eq.({\ref{eq.for QNM to 8}}), 
%------------------------------polar coordinate.--------------------------------------%
\begin{equation}
(\omega R)^8 = r e^{i\theta},
\label{eight}
\end{equation}
%----------------------------------------------------------------------------------%
 we obtain two real equations,
%--------------------------------two equation-----------------------------------%
\begin{eqnarray}
0 = && 1 + \frac{\pi^2}{2^9}r \cos \theta \nonumber \\
 &&  - \frac{\pi r}{2^{10}}\left[
(\log r+8\log\frac{e^{\gamma}}{2})\sin\theta + \theta \cos \theta
\right]. 
\label{real}
\end{eqnarray}
\begin{eqnarray}
0 = 2\pi \sin \theta+\left[
 (\log r+8\log \frac{e^{\gamma}}{2})\cos\theta -\theta \sin\theta
\right]. 
\label{imaginary}
\end{eqnarray}
%----------------------------------------------------------------------------------%
From Eq.(\ref{imaginary}), $r$ is expressed in terms of $\theta$ as,
%----------------------------------for r-------------------------------------------%
\begin{equation}
r=2^{8}e^{-8\gamma}\exp[(\theta-{2\pi }) \tan\theta ].
\label{r=}
\end{equation}
%----------------------------------------------------------------------------------%
So, we get the equation for $\theta$ from Eq.(\ref{real}) and Eq.(\ref{r=}),
%--------------------------------eq. for theta------------------------------------%
\begin{equation}
1-\frac{ \pi}{4 e^{8\gamma}}e^{(\theta-{2\pi})\tan \theta}
\frac{ \theta-{2\pi} }{\cos(\theta-{2\pi})}=0.
\label{eq for theta}
\end{equation}
%----------------------------------------------------------------------------------%
Since, $r$ must satisfy $r \ll 1$, we can see from Eq.(\ref{real})
 that $\theta$ must satisfy $|\theta| \gg 1$. 
There are two cases for the solution of Eq.(\ref{r=}) and Eq.(\ref{eq
 for theta}) as follows, 
%----------------------------------------------------------------------------%
\begin{eqnarray}
&(i)& ~~ \theta > 2\pi,~ \tan \theta<0,~ \cos(\theta-2\pi)>0 \nonumber\\
& \Leftrightarrow&~ -\frac{\pi}{2}+2n\pi < \theta < 2n\pi,~ n\geq 2 \\
&(ii)& ~~\theta< 2\pi,~ \tan\theta>0,~ \cos(\theta-2\pi)<0 \nonumber\\
&\Leftrightarrow&~ -(2n-1)\pi<\theta<-(2n-1)\pi+\frac{\pi}{2},~n\geq 0. \nonumber \\ 
\end{eqnarray}
%-------------------------------------------------------------------------------%
where $n$ is an integer.
The condition $|\theta| \gg 1$ is satisfied when $n \gg 1$.  
Let us consider the case (i). We put $\theta = 2n\pi - \Delta_n$ 
and $0< \Delta_n<\pi/2$.
Then Eq.(\ref{eq for theta}) can be written as follows,
%----------------------------------------------------------------------------------%
\begin{eqnarray}
1-\frac{ \pi}{4 e^{8\gamma}}e^{-2n\pi\tan \Delta_n}
\frac{{2n\pi} }{\cos \Delta_n} \simeq 0.
\end{eqnarray}
%-----------------------------------------------------------------------------------%
$\Delta_n$ must be much less than unity 
in order that the second term of the above
equation cancels the first term. It follows that,
%----------------------------------------------------------------------------------%
\begin{eqnarray}
 1-\frac{ \pi}{4 e^{8\gamma}}e^{-2n\pi\Delta_n}
2n\pi  \simeq 0, \\
\Rightarrow
\Delta_n \simeq \frac{1}{2n\pi}(\log n+\log \frac{\pi^2}{2e^{8\gamma}}). 
\label{Delta_n}
\end{eqnarray}
%----------------------------------------------------------------------------------%
In the case (ii), we put $\theta = -(2n+1)\pi+\Delta_n$.
After a similar analysis, we have the same result Eq.(\ref{Delta_n})
for $\Delta_n$.
This $\Delta_n$ are very small number as long as $n\gg 1$,
so that the solution of Eq.(\ref{eq.for QNM to 8}) 
is just below the real axis
of complex $(\omega R)^8$ plane.
The radial parameter $r$ can be written as,
\begin{eqnarray}
r=\frac{2^9}{\pi^2 n}+{\mathcal{O}}
\left(\frac{\log n}{n^2}\right)
\end{eqnarray}
The QNMs frequency are $R^{-1}$ times eighth roots of $re^{i\theta}$.

\section{The wave scattering of D3-brane in space-time approach}

In this section, we consider the case of a small number of the
D3-brane in string theory, so that the description of the system
based on the field theory with Dirac-Born-Infeld action
is effective one rather than that in terms of the supergravity
which has been used in the previous section.
For simplicity, we treat one D3-brane in flat space-time 
and field theory with DBI action on it.
In the case of the dilaton absorption by D3-brane\cite{K,GHKK,GH},
the evaluation of the cross section is based on 
the world-volume interpretation, in which bulk dilaton field 
is recognized as a source of the world-volume fluctuation. 
In other words, it is assumed that quantum fluctuation of the bulk fields 
decouples and the dynamics is strictly on the brane.  

As explained in the Appendix A, QNMs are determined by the condition 
(\ref{general cond of QNM})
which states no incident wave in the scattering of the dilaton field 
in the black hole spacetime, and the reflective amplitude is needed to
evaluate QNMs frequency.
The world-volume approach \cite{K,GHKK,GH} do not give reflective
amplitude which is concerned to bulk propagation. 
Thus the world-volume approach\cite{K,GHKK,GH} is not suitable 
for investigation of QNMs.
Since the above definition of QNMs (\ref{general cond of QNM})
based on the scattering processes of the dilaton, we have to 
take account of its bulk propagation at least.         
In order to keep close similarity to the evaluation of QNMs 
in previous section based on the classical supergravity, 
we throw a dilaton into a D3-brane from spatial infinity, 
which is  treated  as a wave in field theory 
such as R.P.Feynman's space-time approach
\cite{Feynman 1,Feynman 2,Feynman 3,JJS}.
Incident dilaton propagates in the bulk space and 
eventually interacts with a D3-brane.
By perturbatively solving the wave equation for the dilaton, 
we can obtain its reflective amplitude.
Another reason for using this space-time approach is that we
bear its waveform in mind,
which is one of the other important aspects of QNMs. 
The calculation based on CFT \cite{BSS} might be enough to 
evaluate the frequencies of QNMs which are given by positions of poles 
of the retarded CFT correlator. However our approach can tell us actual
wave forms observed at infinity. Thus it is possible to 
discuss relation between two pictures of black hole spacetime, i.e.
the supergravity and the D-brane description, from the viewpoint not only 
of the frequencies of QNMs but also their wave forms. 
In the appendix C, for a proof of the validity of our calculation based on 
the Feynman's spacetime approach, the dilaton 
absorption cross section is evaluated by the formalism which developed in the 
present paper. And we show that our result agrees in the lowest order
with that obtained by the world volume approach\cite{K,GHKK,GH}.  

For a low-energy dilaton, DBI action can be expanded in Einstein frame
as follows\cite{GHKK},
%---------------------------DBI action------------------------------------%
\begin{widetext}
\begin{eqnarray}
 S_{{\rm{DBI}}}  
 &=&
-T_3\int d^4x\sqrt{-det\left(
G_{\mu\nu}^E+\frac{e^{-\frac{\phi}{2}}}{\sqrt{T_3}}
F_{\mu\nu}
\right)} \nonumber \\
 &=&
 \int d^4 x \left(
\frac{1}{4}F^2-\frac{\sqrt{2}}{4}\kappa_{10}
\phi F^2+\frac{1}{8T_3}(F^4-\frac{1}{2}(F^2)^2)
\right)  + \cdots ,
\end{eqnarray}
\end{widetext}
%-------------------------------------------------------------------------------%
where\
$ F^n=F^{\mu 1}_{\ \mu 2}\cdots F^{\mu{\small{n}}}_{\ \mu 1}$
and $T_3$ is the tension of the D3-brane.
The term $F_{\mu\nu}$ is field strength of the gauge field on the
3-brane describing its longitudinal dynamics, 
while its transverse oscillations do not couple with dilaton.
For $N=1$ case, gauge fields on the D3-brane are photon, and  
we take Coulomb gauge for it on the D3-brane. 
Then the  action for the dilaton in the bulk and the gauge fields on the brane 
is given by
%----------------------------action--------------------------------------------%
\begin{widetext}
\begin{eqnarray}
S_{{\rm{eff}}} = 
 -  \frac{1}{2}
\int d^{10}x 
\partial_{A}\phi\partial^{A}\phi 
 - \frac{1}{2}\int d^{10}x \  \delta^{(6)}(x)
\left[
(1-\sqrt{2}\kappa_{10}\phi)
\partial_{\mu}A_{i}\partial^{\mu}A^{i} 
+\frac{1}{2T_3}(\epsilon^{ijk} \dot{A}_i \partial_j A_k)
(\epsilon^{lmn} \dot{A}_l \partial_m A_n)
\right],
\end{eqnarray}
\end{widetext}
%-----------------------------------------------------------------------------%
where $i=1,2,3$ and $\dot{A} = \partial_0 A$.
For a low energy dilaton, the scattering process is dominated by s-wave 
so that we can put its form as $\phi=\phi(t,r)$ where  $r^2=x_4^2+\cdots+x_9^2$.
The evolution of the dilaton in the bulk and the gauge filed on the brane 
are expressed by the wave equations,   
%----------------------------------------------------------------------------%
\begin{eqnarray}
(-\partial_{t}^2+\partial_{r}^2-\frac{15}{4r^2})\varphi & = & - \delta (r)
\frac{\kappa_{10}}{\sqrt{2}\pi^3r^{5/2}}
\partial_{\mu}A_i\partial^{\mu}A^i, \\
\partial_{\mu}
\partial^{\mu}A^i &  = & \sqrt{2} \kappa_{10}
\partial_{\mu}(\phi\partial^{\mu}A^i) \nonumber \\
&& +  \frac{1}{T_3}\partial_0\left[
(\epsilon^{lmn} \dot{A}_l \partial_m A_n)\epsilon^{ijk}\partial_j A_k
\right] \nonumber \\
&& +\frac{1}{T_3}\partial_a
\left[
(\epsilon^{lmn} \dot{A}_l \partial_m A_n)\epsilon^{bai}\dot{A}_b
\right], \nonumber \\
\end{eqnarray}
%------------------------------------------------------------------------------%
where we change the variable as $\varphi=r^{5/2}\phi$.
We can solve these wave equations by means of  perturbation with respect to 
interaction with gauge fields, i.e, the coupling constant $\kappa_{10}$,
and four point interaction of gauge fields, whose coupling constant is  
$\frac{1}{T_3}$. At first, these waves are formally expanded in terms of 
the two coupling constants, 
%--------------------------------------perturbation------------------------%
\begin{eqnarray}
\varphi &=& \varphi_{(0,0)}+\kappa_{10}\varphi_{(1,0)}
+\frac{\kappa_{10}}{T_3}\varphi_{(1,1)} \nonumber \\
&&\qquad\qquad\quad   +\kappa^2_{10}\varphi_{(2,0)}
+\frac{\kappa_{10}^2}{T_3}\varphi_{(2,1)}+ \cdots, \\
A_i &=& A_{(0,0)}^i+\kappa_{10}A_{(1,0)}^i+\frac{1}{T_3}A_{(0,1)}
+\frac{\kappa_{10}}{T_3}A^i_{(1,1)}+ \cdots.\nonumber \\
\end{eqnarray}
%-------------------------------------------------------------------------------%
We are interested only in response of the D3-brane to the incident 
dilaton wave. So we omit terms $ A_{(0,1)} $ and $\phi_{(1,0)}$ in 
the above expansions, since these terms represent excitation 
which exists before arrival of the injected dilaton at 3-brane. 

At $(0,0)$-order, we obtain wave equations  
%---------------------------------(0,0)eq-----------------------------------------%
\begin{eqnarray}
(-\partial_t^2+\partial_r^2-\frac{15}{4r^2})\varphi_{(0,0)} &=& 0, \\
\partial_{\mu}\partial^{\mu}A^i_{(0,0)} &=& 0 .
\label{(0,0)eq}
\end{eqnarray}
%------------------------------------------------------------------------------%
which express that   
the dilaton wave does not interact with the D3-brane at the origin 
$r=0$, and pass through it.
So, regularity of the dilaton at the origin requires that 
the wave is reflected with the same amplitude as 
that of incidental one. 
The solutions in the lowest order are given by
%-------------------------------------(0,0)solution------------------------------%
\begin{eqnarray}
\varphi_{(0,0)} & = & C e^{-i\omega t}\sqrt{r}\left[
H_2^{(1)}(\omega r)+H_2^{(2)}(\omega r)
\right], \\
A_{(0,0)}^i & = &  \int \frac{d^3k}{(2\pi)^3}\sqrt{\frac{V_3}
{2k_0}}({\mathbf{\epsilon}}_{\mathbf{k}}^{\alpha })^i
e^{i{\mathbf{k}}\cdot{\mathbf{x}}} \nonumber \\
&&  \times 
 \left[
e^{ik_0t}\sum_{\alpha =1}^{2} C_\alpha^* +
e^{-ik_0t}\sum_{\alpha=1}^{2}C_\alpha
\right].
\end{eqnarray}
%-----------------------------------------------------------------------------%
where, $k_0=|{\mathbf{k}}|$,
 $({\mathbf{\epsilon}}_{\mathbf{k}}^{\alpha})^i$
 is polarization vector, and the complex number $C_\alpha$  
, which satisfies
 $|C_\alpha |^2=1$, represents a degree of freedom for the phase 
of the initial configuration of the gauge field. 
We prepare the wave function in the lowest order, $A^i_{(0,0)}$,
to express its vacuum fluctuation of the photon so that
its normalization  is taken to 
be as same as that in the corresponding quantum field theory.

At $(1,0)$-order, the wave equations for the gauge field is
%-------------------------------------(1,0)eq-------------------------------------%
\begin{eqnarray}
\partial_{\mu}\partial^{\mu}A^i_{(1,0)} &=& 
-\sqrt{2}\dot{\varphi}_{(0,0)}\dot{A}_{(0,0)}^i.
\label{(1,0)eq}
\end{eqnarray}
%---------------------------------------------------------------------------------%
Here we choose a negative energy photon $A_{(0,0)}$ in Eq.(\ref{(1,0)eq}).
Thus this equation shows that 
the positive energy photon $A^i_{(1,0)}$ is created from 
the positive energy dilaton $\varphi_{(0,0)}$ and the negative energy 
photon $A^i_{(0,0)}$. 
Because, in standard prescription of quantum field theories, a negative 
energy wave function is interpreted as an anti-particle which 
moves opposite direction, 
Eq.(\ref{(1,0)eq}) describes a dilaton annihilation on the D3-brane and 
pair creation of photons as the calculation based on CFT in \cite{K}.
In order to construct the solution of Eq.(\ref{(1,0)eq}), we use the 
retarded Green's function for 4-dimensional d'Alembertian,
%---------------------------------Green's func. for 4-dim d'Alembertian------------% 
\begin{equation}
G_A(t,{\mathbf{x}};t',{\mathbf{x}}') = - 
\frac{\delta(t-t'-|{\mathbf{x}}-{\mathbf{x}}'|)}
{4\pi |{\mathbf{x}}-{\mathbf{x}}'|},
\label{Gf4}
\end{equation}
satisfying
\begin{equation}
\partial_{\mu}\partial^{\mu}G_A(x;x') = \delta^{(4)}(x-x'),
\label{Green for A}
\end{equation}
%-----------------------------------------------------------------------------------%
which leads to created photon on the D3-brane,
%------------------------------------(1,0)sol-------------------------------------%
\begin{eqnarray}
A_{(1,0)}^i = && \frac{\sqrt{2}C\omega^3}{8}\int \frac{d^3k}{(2\pi)^3}
\sqrt{\frac{V_3}{2k_0}}
e^{-i(\omega-k_0)t}e^{ik_ix^i} \nonumber \\
&&\times \left(
\frac{1}{\omega}+\frac{1}{2k_0-\omega-i\epsilon}
\right)
\sum_{\alpha=1}^2C_\alpha^*(\epsilon_{k}^{\alpha})^i.
\label{(1,0)sol}
\end{eqnarray}
%---------------------------------------------------------------------------------%

The wave equation for the dilaton at (2,0)-order is given by
%--------------------------------------(2,0)eq-----------------------------------%
\begin{equation}
(-\partial_t^2+\partial_r^2-\frac{15}{4r^2})\varphi_{(2,0)} = 
-\frac{2 \delta(r)}{\sqrt{2}\pi^3r^{\small {5/2}}}
\partial_{\mu}A_{(0,0)}^i\partial^{\mu}A_{(1,0)}^i.
\label{(2,0)eq}
\end{equation}
%-----------------------------------------------------------------------------%
The homogeneous part of the solution for the above equation
should be zero, since there is no incident dilaton at (2,0)-order. 
And the Eq.(\ref{(2,0)eq}) represents the process that 
the excited gauge field $A_{(1,0)}^i$ and positive energy 
photon $A_{(0,0)}^i$ annihilate to emit a dilaton wave away from the D3-brane.
We can construct  retarded Green's function for dilaton, 
%----------------------------------Green's function for dilaton---------------%
\begin{eqnarray}
&& G_R(t,r ; t',r') = \frac{\pi}{2i}\theta (t-t')\int\frac{d\omega}{2\pi}
e^{-i\omega (t-t')} \nonumber \\
&& \times \left[
\theta(r-r')H_2^{(1)}(\omega r)J_2(\omega r')\sqrt{rr'}+
(r\leftrightarrow r')
\right],
\end{eqnarray}
%-----------------------------------------------------------------------------%
which satisfies the equation,
%----------------------------------------------------------------------------%
\begin{equation}
(-\partial_t^2+\partial_r^2-\frac{15}{4r^2})G_R(t,r ; t',r')
 = \delta (t-t') \delta (r-r').
\label{Green for d}
\end{equation}
%---------------------------------------------------------------------------%
The explicit expression (\ref{Green for d}) shows that this Green's function
is regular at the origin $r=0$ and obeys the outgoing condition at the
spatial infinity $r \to \infty$.
%%%%%%%%%%%%%%%%%%%%%%%%%%%%%%%%%%%%%%%%%%%%%%%%%%%%%%%%%%%%%%%%%%%%%%%%%%
\begin{figure}[tbh]
\vspace{0.5cm}
\includegraphics[width=4.5cm,height=2cm]{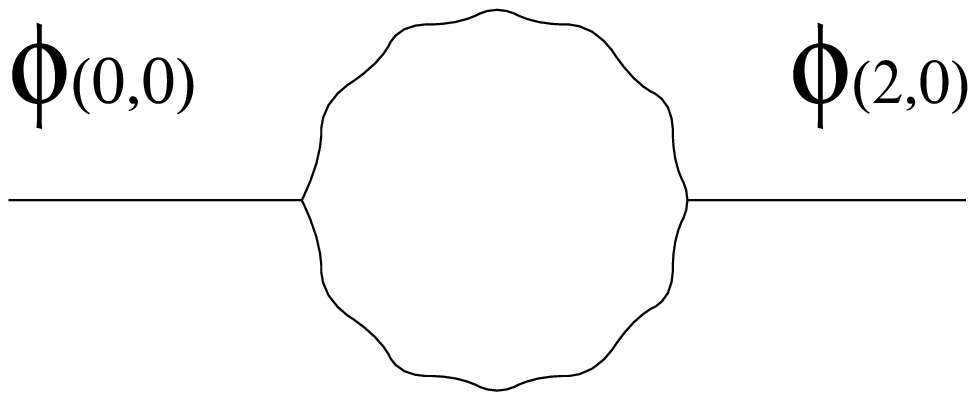}% Here is how to import EPS art
\caption{\label{fig:1-loop}  The 1-loop self energy for the dilaton. A straight and 
wavy lines represent propagating dilaton and photon,respectively }
\end{figure}
%%%%%%%%%%%%%%%%%%%%%%%%%%%%%%%%%%%%%%%%%%%%%%%%%%%%%%%%%%%%%%%%%%%%%%%%%%%
Then the $(2,0)$ dilaton which is integrated over the D3-brane is given by
%------------------------------------(2,0)sol---------------------------------%
\begin{eqnarray}
\bar{\varphi}_{(2,0)} &=& \frac{1}{V_3} 
\int_{V_3} d^3x~~ \varphi_{(2,0)} \nonumber \\ 
                &=& \displaystyle 
 -\frac{C\  \omega^8}{32\cdot 64\pi^3}e^{-i\omega t}
H_2^{(1)}(\omega r)\sqrt{r} \nonumber \\
&& \times \left(1+\frac{i}{\pi}\log\frac{\omega}{\Lambda}\right),%\\
%\nonumber
\label{bar}
\end{eqnarray}
%-----------------------------------------------------------------------------%
where $\Lambda$ is a cut-off factor.
The Eq.(\ref{(2,0)eq}) tells us that $\bar{\varphi}_{(2,0)}$ is created 
through annihilation of two photons, $A^i_{(0,0)}$ and $A^i_{(1,0)}$,
which are described  by Eq.(\ref{(1,0)eq}). As discussed before, it 
can be interpreted that two photons are emerge via the pair creation
 process. Thus in the framework of quantum field theories, 
the external line corrected by the 1-loop self energy Feynman diagram 
depicted in Fig. \ref{fig:1-loop}                              
would correspond to the dilaton wave function $\bar{\varphi}_{(2,0)}$.
%so that we have introduced a symmetric factor $1/2$ in 
%Eq.(\ref{bar}).
%%%
We note that the field $A_{(0,0)}$ appeared in Eq.(\ref{(1,0)eq}) and that in
Eq.(\ref{(2,0)eq}) are identical ones as shown in Fig.\ref{fig:1-loop},
although we interpret these two photons as particle-antiparticle pair. 
Thus, in the evaluation of Eq.(\ref{bar}), momentum integrations are doubly counted
so that we have introduced a symmetric factor $1/2$. 
The divergent contribution to the $(2,0)$-order dilaton should be removed by
the renormalization procedure in which the cut-off $\Lambda$ is 
replaced by a scale for renormalization $R$, i.e. the characteristic length of
the black hole. 

\begin{figure}[tbh]
\vspace{0.2cm}
\includegraphics[width=4.5cm,height=1.5cm]{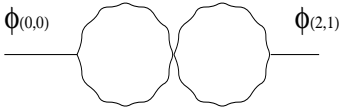}% Here is how to import EPS art
\caption{\label{fig:2-loop}  The 2-loop self energy for the dilaton. A straight and 
wavy lines represent propagating dilaton and photon,respectively }
\end{figure}

At $(1,1)$-order, the equation for $\varphi_{(1,1)}$ have 
no source terms and is not created.
The equation for $A_{(1,1)}^i$ is given by
%--------------------------(1,1)eq-------------------------------------------%
\begin{widetext}
\begin{eqnarray}
\partial_{\mu}\partial^{\mu} A_{(1,1)}^i = \partial_0
\left[
\dot{{\mathbf{A}}}_{(1,0)}\cdot(\bigtriangledown \times {\mathbf{A}}_{(0,0)})
\epsilon^{ij}_{\ \ k}\partial_j A^k_{(0,0)}
\right]
+\partial_a\left[
\dot{{\mathbf{A}}}_{(1,0)}\cdot(\bigtriangledown \times {\mathbf{A}}_{(0,0)})
\epsilon^{bai}\dot{A}_{(0,0)b}
\right] \nonumber \\
+
\partial_0
\left[
\dot{{\mathbf{A}}}_{(0,0)}\cdot(\bigtriangledown \times {\mathbf{A}}_{(1,0)})
\epsilon^{ij}_{\ \ k}\partial_j A^k_{(0,0)}
\right]
+\partial_a\left[
\dot{{\mathbf{A}}}_{(0,0)}\cdot(\bigtriangledown \times {\mathbf{A}}_{(1,0)})
\epsilon^{bai}\dot{A}_{(0,0)b}
\right] \nonumber \\
+
\partial_0
\left[
\dot{{\mathbf{A}}}_{(0,0)}\cdot(\bigtriangledown \times {\mathbf{A}}_{(0,0)})
\epsilon^{ij}_{\ \ k}\partial_j A^k_{(1,0)}
\right]
+\partial_a\left[
\dot{{\mathbf{A}}}_{(0,0)}\cdot(\bigtriangledown \times {\mathbf{A}}_{(0,0)})
\epsilon^{bai}\dot{A}_{(1,0)b}
\right].
\label{(1,1)eq}
\end{eqnarray}
\end{widetext}
%----------------------------------------------------------------------------%
This equation describes four point interaction of photons.
Similarly as the case of $(1,0)$-order photon (\ref{(1,0)eq}),
one of the $(0,0)$-photon $A^i_{(0,0)}$ is chosen to the negative energy 
photon in the right hand side so that Eq.(\ref{(1,1)eq}) 
can be interpreted 
as two photon scattering. 
These scattered $(1,1)$-order photon $A_{(1,1)}$ and $(0,0)$-order photon
$A^i_{(0,0)}$ are 
annihilated to produce  $(2,1)$-order dilaton as follows,
%------------------------------(2,1)eq------------------------------------------%
\begin{eqnarray}
(-\partial_t^2+\partial_r^2-\frac{15}{4r^2})  \varphi_{(2,1)}=\hspace{8em} 
\nonumber \\
- \delta (r) \frac{\sqrt{2}}{\pi^3r^{\small{5/2}}}
\partial_{\mu}A_{(0,0)i}\partial^{\mu}A_{(1,1)}^i.
\label{(2,1)eq}
\end{eqnarray}
%---------------------------------------------------------------------------%
The source term of this equation is integrated over the D3-brane to give
%------------------------------Source of(2,1)--------------------------------%
\begin{eqnarray}
&& \frac{1}{V_3}\int d^3 x
\partial_{\mu}A_{(0,0)i}\partial^{\mu}A_{(1,1)}^i
= \nonumber \\
&& \frac{C \omega^{10}}{3\cdot 2^{11}\sqrt{2}\pi^4}e^{-i\omega t}
\left(
\pi^2+2\pi i \log\frac{\omega}{\Lambda}-(\log\frac{\omega}{\Lambda})^2
\right).~~~
\end{eqnarray}
%---------------------------------------------------------------------------%
Using  Green's function for a dilaton Eq.(\ref{Green for d}),
we obtain the solution of Eq.(\ref{(2,1)eq}) integrated over the 
D3-brane,
%---------------------------(2,1)sol--------------------------------------------%
\begin{eqnarray}
\bar{\varphi}_{(2,1)}& & 
= \frac{1}{V_3} \int_{V_3} d^3 x ~\varphi_{(2,1)} \nonumber \\
=&&   -\int dt' dr' G_R 
\frac{\sqrt{2}\delta(r')}{\pi^3 {r'}^{5/2}}
\frac{1}{V_3}\int d^3 x
\partial_{\mu}A_{(0,0)i}\partial^{\mu}A_{(1,1)}^i \nonumber \\
 =&& 
C\frac{\omega^{12}e^{-i \omega t}}
{3\cdot 2^{14}\pi^4} 
 H_2^{(1)}(\omega r)\sqrt{r} \nonumber \\
&& \times
\left(
i-\frac{2}{\pi}\log\frac{\omega}{\Lambda}
-\frac{i}{\pi^2}(\log\frac{\omega}{\Lambda})^2
\right),%\nonumber \\
\end{eqnarray}
%----------------------------------------------------------------------------------%
where we again introduce a symmetric factor 1/4,
since the double counting in the momentum integration, which emerged in the
evaluation of Eq.(\ref{bar}), 
occurs twofold in this case. 
%because $A_{(0,0)}$ in Eq.(\ref{(1,0)eq}) and one of it in Eq.(\ref{(1,1)eq})
%are field theoretically the same fields as in Fig.\ref{fig:2-loop}
% and their momentums is double counted 
%and so is  another $A_{(0,0)}$ in Eq.(\ref{(1,1)eq}) and in Eq.(\ref{(2,1)eq}).
The cut-off 
$\Lambda$ will be replaced by the inverse of the characteristic length
of the black hole $R^{-1}$ by the procedure of renormalization.
The obtained wave function $\bar{\varphi}_{(2,1)}$ 
correspond to the external line corrected by the 2-loop Feynman
diagram depicted in Fig.\ref{fig:2-loop}.
%%%%%%%%%%%%%%%%%%%%%%%%%%%%%%%%%%%%%%%%%%%%%%%%%%%%%%%%%%%%%%%%%%%%%%%%%%
%%%%%%%%%%%%%%%%%%%%%%%%%%%%%%%%%%%%%%%%%%%%%%%%%%%%%%%%%%%%%%%%%%%%%%%%%%%

Assembly of the calculated dilaton waves at several orders leads to
%---------------------------total phi----------------------------------------------%
 \begin{eqnarray}
\bar{\varphi} &=& \bar{\varphi}_{(0,0)}+\kappa^2_{10}\bar{\varphi}_{(2,0)}
+\frac{\kappa_{10}^2}{T_3}\bar{\varphi}_{(2,1)}+ \cdots \\
 &=& C e^{-i\omega t} \sqrt{r}
 \left[
 {\mathcal{R}}H^{(1)}_2(\omega r)+
 H^{(2)}_2(\omega r)
 \right] \\
 &\rightarrow&
 C e^{-i\omega t} \sqrt{\frac{2}{\pi\omega}}
 \left[
 {\mathcal{R}}e^{i(\omega r-\frac{5}{4}\pi)}+
 e^{-i(\omega r-\frac{5}{4}\pi)}
 \right],
\label{total phi}
\end{eqnarray}
%----------------------------------------------------------------------------------------%
where in the last line, 
we take $r\to \infty$ limit and  ${\mathcal{R}}$ is the reflective amplitude,
%-----------------------------------------------------------------------------------------%
\begin{eqnarray}
&& {\mathcal{R}} =  
1-\frac{\kappa_{10}^2\omega^8}{2^{11}\pi^3}
\left(
1+\frac{i}{\pi}\log\omega R
\right) \nonumber \\
&& +\frac{\kappa_{10}^2}{T_3}\frac{\omega^{12}}{3\cdot2^{18}\pi^4}
\left(
i-\frac{2}{\pi}\log\omega R
-\frac{i}{\pi^2}(\log\omega R)^2
\right)
+\cdots \nonumber \\
\label{reflection}
\end{eqnarray}
%-------------------------------------------------------------------------------------%
Here, we recall the relation between $\kappa_{10}$, $T_3$ and $R$ 
%-------------------------------parameter relation----------------------------------%
\begin{equation}
\kappa_{10}=\frac{\sqrt{\pi}}{T_3},
\qquad
R^4 = \frac{\kappa_{10}}{2\pi^{5/2}},
\end{equation}
%------------------------------------------------------------------------------%
where we set $N=1$.
Thus the condition of QNM discussed in the appendix A is given by
%-------------------------------------------------------------------------------%
\begin{eqnarray}
0  & = &  \frac{\mathcal{I}}{\mathcal{R}}e^{i\beta'} \nonumber \\
& = &  1  + 
\frac{\pi^2}{512}(\omega R)^8
\left(
1+\frac{i}{\pi}\log\omega R
\right) \nonumber \\
&& +  \frac{\pi^2}{512\cdot 12}(\omega R)^{12}
\left(
\log\omega R
+\frac{i}{2\pi}(\log\omega R)^2
\right) \nonumber \\
&& +{\mathcal{O}((\omega R)^{16})}, 
\label{eq for QNM}
\end{eqnarray}
%------------------------------------------------------------------------------%
where $\beta'$ is a phase factor,
%---------------------------------------------------------------------------%
\begin{equation}
\beta' = \frac{\pi^3}{512\cdot 24}(\omega R)^{12}.
\end{equation}
%-----------------------------------------------------------------------------%
Eq.(\ref{eq for QNM}) up to $(\omega R)^8$ is very similar with 
Eq.(\ref{eq.for QNM to 8}), and the solution has the qualitatively 
equivalent behavior.

\section{Discussion}
 
We have evaluated quasi-normal modes (QNMs) of the D3-brane black hole 
by two different approach, 
the one is based on the black hole picture in terms of  
the black 3-brane solution of the low-energy supergravity action and 
the other the D3-brane  described by the field theory 
with the Dirac-Born-Infeld action. 
We have shown that the two different method derived qualitatively 
same condition for the QNMs for the low-energy region. 
As described in appendix C, the evaluations of absorption cross section 
shows similar agreement. 
In both cases, difference, which appears in logarithmic terms,
seems to imply that we must take proper account of 
the non-abelian nature of the theory\cite{GHKK,GH,CGKT}.

The condition for QNMs (\ref{general cond of QNM}) has been obtained 
as Eq.(\ref{eq for QNM in SUGRA}) and Eq.(\ref{eq for QNM}) 
in terms of an expansion with respect to $\omega R$.
Up to the order $(\omega R)^8$, we can express the condition of QNM
as follows,
%---------------------------------------------------------------------------%
\begin{eqnarray}  
\frac{{\mathcal{I}}}{{\mathcal{R}}} \sim 
1+ i\,\Sigma = 0,
\label{1+Sigma} 
\end{eqnarray}
%-----------------------------------------------------------------------%
where $\sim$ means equality up to phase factors and
%------------------------------------------------------------------------%
\begin{eqnarray}
\Sigma =
\frac{\pi^2}{512}(\omega R)^8
\left(\frac{\alpha}{\pi}\log\frac{\omega \bar{R}}{2} - i \right). 
\label{Sigma}
\end{eqnarray}
%------------------------------------------------------------------------%
In the field theoretical approach as in section 3,
 $\Sigma$, which is just the
coefficient of $\phi_{(2,0)}$, 
corresponds to the 1-loop self energy of dilaton. 
The procedure of the spacetime approach gives 
the dilaton wave function as (\ref{total phi}), and 
we can read off reflective amplitude
(\ref{reflection}) which is expressed as
\begin{equation}
\frac{\mathcal{R}}{\mathcal{I}} \sim 1- i\,\Sigma 
\end{equation}
up to the order $(\omega R)^8$. In our calculation 
this reflective amplitude is perturbatively inverted 
to give Eq.(\ref{1+Sigma}). 
However, in the viewpoint of evaluation based on field theories, 
the above inversion can be interpreted as 
the summation of the self-energy diagrams
%%-----------------------------------------------------------------------%
\begin{equation}
\frac{\mathcal{R}}{\mathcal{I}} \sim 1 + (-i\Sigma) + (-i\Sigma)^2 + \cdots 
\sim \frac{1}{1 + i \Sigma}.
\label{sum of self energy}
\end{equation}
%-----------------------------------------------------------------------%
Thus QNMs can be understood as poles of the scattered propagator
in the field theory on D3-brane. 
This is realized in the calculation of AdS/CFT correspondence
in the BTZ black hole case in \cite{BSS}

\begin{acknowledgments}
We would like to thank T.Uematsu and S.Yamaguchi for useful discussions. 
\end{acknowledgments}

\appendix

\section{A brief review of QNM}

In this appendix, we give a brief review of Quasi-normal modes (QNMs)
which represent the relaxation process in a black hole spacetime.
If one perturbes the black hole spacetime, then 
gravitational wave is emitted to both directions. One part of the wave 
propagates to spatial infinity and other falls into black hole horizon.
So, it is natural to specify the boundary condition for
such waves as follows, ingoing at horizon and
outgoing at spatial infinity.

Let us consider a minimal scalar on the black holes background.
It satisfies the wave equation of following forms 
after some change of variables\cite{KS},
%----------------------------wave equation---------------------------------%
\begin{equation}
(-\partial_t^2+\partial_{r_*}^2-V(r_*))\phi=0
\label{wv}
\end{equation}
%--------------------------------------------------------------------------%
where $r_*$ is the ``tortoise'' radial coordinate.
In the case of the Schwarzschild black hole with the horizon radius
$r = r_{h}$, the coordinate $r_*$ is related to 
the usual radial coordinate $r$ as
$$
r_* = r +  r_{h} \ln \Bigl(\frac{r-r_h}{r_h}\Bigr),
$$
which spans the region outside of the horizon of the black hole.
It follows that $r_* \rightarrow -\infty$ represents approaching 
the horizon ($r=r_h$) and  $r_* \rightarrow \infty$ the spatial 
infinity ($r \rightarrow \infty$), respectively.  
And the potential 
$V(r_*)$, which has the information of the curvature of the spacetime, 
is  assumed to have the asymptotic behavior, 
$V(r_*) \to 0$ as $r_* \to -\infty$ (to horizon) 
and $r_* \to \infty$ (to spatial infinity).
Under this assumption,
the solution of wave equation (\ref{wv})
with a frequency $\omega$ can be expressed as 
%------------------------------general solution-----------------------------%
\begin{equation}
\phi(t,r_*) = A e^{-i\omega (t-r_*)}+Be^{-i\omega (t+r_*)}
\end{equation}
%---------------------------------------------------------------------------%
in the asymptotic region.
Here let me remind you the boundary conditions, i.e. 
{\it ``QNMs being ingoing at the horizon and 
outgoing at the spatial infinity''} 
are specified by 
\begin{eqnarray}
\phi \propto \left\{
\begin{array}{ll}
 e^{-i\omega (t-r_*)} ,
 & \ (r_* \to \infty) \\
  e^{-i\omega (t+r_*)}, &\   (r_* \to -\infty)
\end{array}\right. 
\label{QNMbc}
\end{eqnarray}

In order to obtain the solution which satisfies the boundary condition
described above, we consider the retarded Green's function
constructed by following prescription.
We prepare two solutions $\phi_1(t,r_*)$ and $ \phi_2(t,r_*)$, which has
asymptotic forms as 
%------------------------------------------------------------------------%
\begin{eqnarray}
\phi_1 \sim \left\{
\begin{array}{ll}
{\mathcal{I}} e^{-i\omega (t+r_*)} +{\mathcal{R}} e^{-i\omega (t-r_*)},
 & \ (r_* \to \infty) \\
 {\mathcal{A}}  e^{-i\omega (t+r_*)}, &\   (r_* \to -\infty)
\end{array}\right. 
\label{1,2 asymptotic}
\end{eqnarray}
\begin{eqnarray}
\phi_2 \sim \left\{
\begin{array}{ll}
{\mathcal{Q}} e^{-i\omega (t-r_*)},&\  (r_* \to \infty) \\
{\mathcal{P}} e^{-i\omega (t-r_*)} + {\mathcal{G}} e^{-i\omega(t+r_*)},
 &\ (r_* \to -\infty) 
\end{array}\right.
%\label{1,2 asymptotic}
\end{eqnarray}
%------------------------------------------------------------------------%
to construct the retarded Green's function
 from them,
%------------------------------------------------------------------------%
\begin{eqnarray}
G_R(t,r_*; t',r_*')=\theta(t-t')\int \frac{d\omega}{2\pi}
e^{-i\omega(t-t')}\frac{1}{\triangle} \nonumber \\
\times \left[
\theta(r_*-r_*')\phi_1(r'_*)\phi_2(r_*)+(r_* \leftrightarrow r_*')
\right],
\label{GreenFunc}
\end{eqnarray}
%--------------------------------------------------------------------------%
where  the Wronskian $\triangle$ can be evaluated as 
$\triangle = 2i\omega {\mathcal{I}}{\mathcal{Q}}$ from the asymptotic 
forms at $r_* \to \infty$,
Using this retarded Green's function, we can obtain the wave $\phi$ 
generated by a source term $S(t,r_*)$
%----------------------------------------------------------------------------%
\begin{equation} 
\phi(t,r_*) = \int dt'dr_*' G_R(t,r_* ; t',r_*') S(t',r_*') ,
\label{solution-by-Int}
\end{equation}
%-----------------------------------------------------------------------------%
which satisfies the inhomogeneous  equation,
%------------------------------------------------------------------------%
\begin{equation}
(-\partial_t^2+\partial_{r_*}^2-V(r_*))\phi = S(t,r_*).
\end{equation}
%--------------------------------------------------------------------------%
%-----------------------------------------------------------------------------%
It is obvious that the solution (\ref{solution-by-Int}) 
satisfies the QNM boundary condition (\ref{QNMbc}).
Taking the limit $r_* \to \infty$ of Eqs. (\ref{solution-by-Int}) and 
(\ref{GreenFunc}), we obtain the asymptotic form of the generated wave 
%----------------------------------------------------------------------------%
\begin{eqnarray}
\phi(t,r_*) \to && \int dt' dr_*' \theta(t-t')
 \int_{\rm C}\frac{d\omega}{2\pi} \nonumber \\
&& \times \frac{e^{-i\omega(t-t')}e^{i\omega r_*}}{2i\omega\,\mathcal{I}}
\phi_1(r_*')S(t',r_*'),
\label{general asymptotic}
\end{eqnarray}
%----------------------------------------------------------------------------%
where contour C must be taken as it enclose lower half of complex
$\omega$ plane for retarded boundary condition.
So, only the poles of lower half plane contribute the $\omega$ integral
in the asymptotic form Eq.(\ref{general asymptotic}) 
and that is the Quasi-normal modes (QNM) 
which are, of course, complex and decaying modes.
In other words, QNM is the poles of retarded Green's function on the 
lower half complex $\omega$ plane. 
The condition of QNM is ${\mathcal{I}}=0$, but only the ratio of
 amplitudes are meaningful in the asymptotic form Eq.(\ref{1,2 asymptotic}),
 so this is equivalent to 
${\mathcal{R}}\to \infty$. Then the more general condition of QNM is,
%-----------------------------------------------------------------------------%
\begin{equation}
{\mathcal{I}}/{\mathcal{R}}=0
\label{general cond of QNM}
\end{equation}
%-----------------------------------------------------------------------------%
The only things we have to do is to solve this equation and choose modes
arising at lower half complex plane.

\section{Explicit expressions}

In this appendix, we summarize explicit form for some results which were 
considered too lengthy to write out in the main text.

The Floquet exponent $\mu$ which satisfies Eq.(\ref{Floquet}) can be 
obtained in the expansion with respect to a power series in $\omega R$,
%--------------------------------------mu---------------------------------%
\begin{widetext}
\begin{equation}
\mu = 1-\frac{i\sqrt{5}}{6}(\frac{\omega R}{2})^4
+\frac{7i}{216\sqrt{5}}(\frac{\omega R}{2})^8
+\frac{11851 i}{62208\sqrt{5}}(\frac{\omega R}{2})^{12}+\cdots.
\end{equation}
%--------------------------------------------------------------------------%
And the coefficient $A_\mu^{(q)}$ in Eq.(\ref{Floquetcoef}) up to 
$(\omega R)^{16}$ are given by 
%-------------------------------------A^(q)--------------------------------%
\begin{eqnarray}
A_{\mu}^{(1)} &=& \frac{1}{3\mu(\mu+1)(\mu+2)}, \nonumber \\
 \ \nonumber \\
A_{\mu}^{(2)} &=& \frac{1}{432}
\left(\frac{1}{\mu}-\frac{11}{\mu+1}-\frac{47}{\mu+2}
\right)
-\frac{1}{144}\frac{1}{\mu+3}-\frac{1}{18}\frac{1}{(\mu+2)^2}
+\frac{5}{36} {\psi}^{(1)} (\mu+2), \nonumber \\
 \ \nonumber \\
\end{eqnarray}
%---------------------------------------------------------------------%
\begin{eqnarray}
A_{\mu}^{(3)} &=&
- \frac{2321}{155520\mu}+\frac{739}{9720(\mu+1)}
    -\frac{5}{216(\mu+2)^3}-\frac{115}{2592(\mu+2)^2} \nonumber \\
&-& \frac{791}{9720(\mu+2)}+\frac{1}{432(\mu+3)^2}
    +\frac{61}{6480(\mu+3)}+\frac{1}{17280(\mu+4)} \nonumber \\
&+& \left(
    \frac{7}{648}+\frac{5}{216\mu}-\frac{5}{108(\mu+1)}
    +\frac{5}{216(\mu+2)}
    \right)
\psi^{(1)}(\mu+2) \nonumber \\
 \ \nonumber \\
A_{\mu}^{(4)} &=& (\frac{164327}{133996800}
-\frac{37\pi^2}{31104})\frac{1}{\mu+1}+\cdots,
\end{eqnarray}
\end{widetext}
%---------------------------------------------------------------------------------------%
where, $\psi^{(1)}$ is the PolyGamma function.

\section{Absorption cross section for D3-brane}

In this appendix, we compare absorption cross section $\sigma$ for 
the D3-brane in various methods for evaluation. 
As showed in \cite{GHKK}, there is a bit difference in $\sigma$ for
 different methods at higher order.

In supergravity, 
the absorption cross section $\sigma$ for the 3-brane 
is given by
%---------------------------------cross section (SUGRA)------------------------%
\begin{eqnarray}
&& \sigma_{\rm sugra} =  \frac{32\pi^2}{\omega^5}
\frac{|{\mathcal{A}}|^2}{|{\mathcal{I}}|^2} \nonumber \\
&& = 
\frac{\pi^4}{8}\omega^3 R^8
\left(
1-\frac{1}{6}(\omega R)^4\log \omega \bar{R}
+{\mathcal{O}}\left((\omega R)^4\right)
\right)\nonumber \\
\end{eqnarray}
%------------------------------------------------------------------------------------%
from Eq.(\ref{asympt}) as in \cite{GH}.
This evaluation is reliable at $N \gg 1$ region where the supergravity 
description of the 3-brane is valid.

We can also calculate the absorption cross section $\sigma$ for the D3-brane 
by means od the space-time approach from Eq.(\ref{total phi}),
%----------------------------sigma (space-time)------------------------------------%
\begin{eqnarray}
&& \sigma_{\rm wave } = \frac{32\pi^2}{\omega^5}
\left(1-
|\frac{{\mathcal{R}}}{{\mathcal{I}}}|^2
\right)\nonumber \\
&& = \frac{\pi^4}{8}\omega^3 R^8
\left(
1+\frac{1}{12}(\omega R)^4\log\frac{\omega}{\Lambda}
+{\mathcal{O}} \left( (\omega R)^4 \right)
\right), \nonumber \\
\end{eqnarray}
%------------------------------------------------------------------------%
where  we set $ N=1 $.

As in \cite{GHKK}, CFT on the D3-brane can also derive 
this quantity, and result is 
%------------------------sigma (Quantum fields theory )------------------------%
\begin{equation}
\sigma_{\rm CFT} = \frac{\pi^4}{8}\omega^3 R^8
\left(
1-\frac{1}{24}(\omega R)^4\log\frac{\omega}{\Lambda}
+ {\mathcal{O}} \left( (\omega R)^4 \right)
\right).
\end{equation}
%--------------------------------------------------------------------------%

These three evaluation of $\sigma$ are agree at the lowest order.

%\newpage %Just because of unusual number of tables stacked at end
%\bibliography{form}% Produces the bibliography via BibTeX.

\begin{thebibliography}{99}


\bibitem{entropy}
J.D.Bekenstein,  
Phys.Rev.D {\bf 5} 1239, (1972) 

\bibitem{AreaTh}
S.W.Hawking
Phys.Rev.Lett. {\bf 26}, 1344 (1971)

\bibitem{HawkingRadiation}
S.W.Hawking
Commun.Math.Phys. {\bf 43} 199, (1975)

\bibitem{GRText}
e.g. R.M.Wald 
``{\it General Relativity}''  
(The University of Chicago Press, Chicago and London, 1984)

\bibitem{QFText}
e.g. N.D.Birrell and P.C.W.Davies 
``{\it Quantum Fields in Curved Space}''  
(Cambridge University Press, Cambridge, 1982)

%% Review of BH entropy in string %%

\bibitem{H}
See e.g,
G.Horowitz
 gr-qc/9604051.

%% original of BH entropy in string %%

\bibitem{SV}
A.Strominger and C.Vafa
Phys. Lett. B {\bf 379}, 99 (1996).

%% original of Hawking rad. in string %%
\bibitem{CM}
C.G.Callan and J.M.Maldacena
Nucl.Phys.{\bf B472},591 (1996).

\bibitem{DM}
S.R.Das and S.D.Mathur
Nucl.Phys.{\bf B 478}, 561 (1996).

%% D3 world volume %%

\bibitem{K}
I. R. Klebanov
Nucl. Phys. {\bf B 496}, 231 (1997).

%% field theory

\bibitem{GHKK}
S. S. Gubser, A. Hashimoto, I. R. Klebanov and M. Krasnitz
Nucl. Phys. {\bf B 526 }, 393 (1998).

%% exact solution

\bibitem{GH}
S. S. Gubser and A. Hashimoto
hep-th/9805140

%% Review of QNM
\bibitem{ChandDet}
Proc. R. Soc. London, {\bf A 344} 441 (1975).

\bibitem{MathematicalBH}
See e.g. S. Chandrasekhar, {\it The Mathematical Theory of Black Holes}
(Clarendon, Oxfprd, 1983)

\bibitem{KS}
See e.g. K.D.Kokkotas and B.G. Schmidt 
Living Reviews in Relaivity
''Quasi-normal modes of Stars and Black Holes'',\\
(http://www.livingreviews.org/Articles/Volume2/1999-2kokkotas/index.html)

%% QNM from AdS/CFT point of view
\bibitem{HH}
G.T.Horowitz and V.E.Hubeny 
Phys.Rev.D {\bf 62} (2000) 024027 

% QNM in AdS background

\bibitem{CM1}
J.S.F. Chan, R.B. Mann
Phys.Rev.D {\bf55} (1997) 7546 

\bibitem{CM2}
J.S.F. Chan, R.B. Mann
 Phys.Rev.D {\bf 59} (1999) 064025

\bibitem{WYA}
B.Wang, C.Lin, E.Abdalla
Phys.Lett.B {\bf 481}  (2000) 79-88

\bibitem{GS}
T.R.Govindarajan, V.Suneeta
Class.Quant.Grav.{\bf 18} (2001) 265-276

\bibitem{ZWA}
Jiong-Ming Zhu, Bin Wang, Elcio Abdalla
Phys.Rev. D {\bf 63} (2001) 124004

\bibitem{B}
Danny Birmingham
Phys.Rev. D {\bf 64} (2001) 064024

\bibitem{CL1}
Vitor Cardoso, Jose' P. S. Lemos
Phys.Rev. D {\bf 64} (2001) 084017

\bibitem{CL2}
Vitor Cardoso, Jose' P. S. Lemos
Class.Quant.Grav.{\bf 18} (2001) 5257-5267

\bibitem{MN}
Ian G Moss, James P Norman
Class.Quant.Grav. {\bf 19} (2002) 2323-2332

\bibitem{WAM}
B. Wang, E. Abdalla, R.B. Mann
Phys.Rev.D {\bf 65} (2002) 084006 

\bibitem{RAK}
R.A.Konoplya
hep-th/0205142, gr-qc/0207028 

\bibitem{AOS}
 Andrei O. Starinets
hep-th/0207133 

%CFT side

\bibitem{BSS}
D.Birmingham, I.Sachs and S.N.Solodukhin
Phys.Rev.Lett {\bf 88}, 151301 (2002).

%% space-time approach

\bibitem{Feynman 1}
R. P. Feynman
Rev. Mod. Phys. {\bf 21}, 425 (1949).

\bibitem{Feynman 2}
R. P. Feynman
Rev. Mod. Phys. {\bf 20}, 367 (1948).

\bibitem{Feynman 3}
R. P. Feynman 
Phys. Rev. {\bf 76}, 749 (1949).

\bibitem{JJS}
J. J. Sakurai
``Advanced Quantum Mechanics''
Addison Wesley Publishing Company 
{\bf Chapter 4-5}



\bibitem{CGKT}
C.G.Callan,S.S.Gubser,I.R.Klebanov,and A.A.Tseytlin, 
Nucl.Phys. {\bf B 489},65 (1997).



\end{thebibliography}

\end{document}